\pdfoutput=1

\documentclass[11pt,twoside,a4paper,cmspaper,final,collab]{cms-tdr}
\def\svnVersion{316892}\def\cmsCernNo{2013-037}\def\cmsCernDate{\today}\def\cmsMessage{Published in the Journal of High Energy Physics as \href{http://dx.doi.org/10.1007/JHEP12(2015)178}{\doi{10.1007/JHEP12(2015)178.}}}

\begin{document}\cmsNoteHeader{HIG-13-035}

\hyphenation{had-ron-i-za-tion}
\hyphenation{cal-or-i-me-ter}
\hyphenation{de-vices}
\RCS$Revision: 314665 $
\RCS$HeadURL: svn+ssh://svn.cern.ch/reps/tdr2/papers/HIG-13-035/trunk/HIG-13-035.tex $
\RCS$Id: HIG-13-035.tex 314665 2015-12-12 20:36:04Z alverson $
\newlength\cmsFigWidth
\ifthenelse{\boolean{cms@external}}{\setlength\cmsFigWidth{0.85\columnwidth}}{\setlength\cmsFigWidth{0.4\textwidth}}
\ifthenelse{\boolean{cms@external}}{\providecommand{\cmsLeft}{top}}{\providecommand{\cmsLeft}{left}}
\ifthenelse{\boolean{cms@external}}{\providecommand{\cmsRight}{bottom}}{\providecommand{\cmsRight}{right}}

\newcolumntype{.}{D{.}{.}{-1}}
\newcolumntype{x}[1]{D{,}{\, \pm \, }{#1}}
\providecommand{\NA}{---\xspace}
\providecommand{\CL}{CL\xspace}

\cmsNoteHeader{HIG-13-035}
\title{Search for a light charged Higgs boson decaying to \texorpdfstring{$\cPqc\cPaqs$}{c-sbar} in \texorpdfstring{$\Pp\Pp$}{pp} collisions at \texorpdfstring{$\sqrt{s} = 8\TeV$}{sqrt(s) = 8 TeV}}

\date{\today}

\abstract{
A search for a light charged Higgs boson, originating from the decay of a top quark and subsequently decaying into a charm quark and a strange antiquark, is presented.
The data used in the analysis correspond to an integrated luminosity of 19.7\fbinv recorded in proton-proton collisions at $\sqrt{s}=8\TeV$ by the CMS experiment at the LHC.
The search is performed in the process ${\ttbar \to \PW^{\pm}\cPqb\PH^{\mp}\cPaqb}$, where the $\PW$ boson decays to a lepton (electron or muon) and a neutrino.
The decays lead to a final state comprising an isolated lepton, at least four jets and large missing transverse energy.
No significant deviation is observed in the data with respect to the standard model predictions, and model-independent
upper limits are set on the branching fraction $\mathcal{B}$($\cPqt\to \PH^{+}\cPqb$), ranging from $1.2$ to $6.5\%$ for a charged Higgs boson with mass between 90 and 160\GeV, under the assumption that $\mathcal{B}$($\PH^{+}\to \cPqc\cPaqs$) = 100\%.
}

\hypersetup{%
pdfauthor={CMS Collaboration},%
pdftitle={Search for a light charged Higgs boson decaying to c-sbar in pp collisions at sqrt(s) = 8 TeV},%
pdfsubject={CMS},%
pdfkeywords={CMS, physics, software, computing}}

\maketitle
\newcommand{\Hpcsbar}{\ensuremath{\PH^{+}\to\cPqc\cPaqs}\xspace}
\newcommand{\tHpb}{\ensuremath{\cPqt\to\PH^{+}\cPqb}\xspace}
\newcommand{\csbar}{\ensuremath{\cPqc\cPaqs}\xspace}

\section{Introduction}
A Higgs boson has recently been discovered by the ATLAS~\cite{Aad:2012tfa} and CMS~\cite{Chatrchyan:2012ufa,Chatrchyan:2013lba} Collaborations with a mass around 125\GeV and properties consistent with those expected from the standard model (SM) within the current experimental uncertainties.
However, precise measurements of the properties of the new boson are needed to identify or exclude differences with respect to the SM predictions.
The mass of the Higgs boson itself is subject to quadratically divergent corrections at high energies~\cite{Quaddiv}. Several extensions beyond the SM (BSM) have been proposed to address such divergences. Supersymmetry~\cite{Golfand:1971iw,Wess:1974tw,Martin:1997ns} is one such model that invokes a symmetry between fundamental fermions and bosons.
The Higgs sector of the so-called minimal supersymmetric standard model (MSSM)~\cite{Fayet:1974pd,Fayet:1977yc} consists of two Higgs doublets, resulting in five physical states: a light and a heavy CP-even $\Ph$ and $\PH$, a CP-odd $\PSA$, and two charged Higgs bosons ${\PHpm}$.
At lowest order, the MSSM Higgs sector can be expressed in terms of two parameters, usually chosen as the mass of the CP-odd boson (${m_{\PSA}}$) and the ratio of the vacuum expectation values of the two Higgs doublets ($\tan\beta$). The generic two-Higgs-doublet model (2HDM), of which the MSSM is a special case, encompasses the following four scenarios.
{\begin{itemize}
\item Type I: All quarks and leptons couple only to the second doublet.
\item Type II: All up-type quarks couple to the second doublet while all down-type quarks and charged leptons couple to the first one.
\item Type X: Both up- and down-type quarks couple to the second doublet and all leptons to the first one.
\item Type Y: The roles of the two doublets are reversed with respect to Type II.
\end{itemize}}

The LEP experiments~\cite{Abbiendi:2013hk} have set a $95\%$ confidence level (\CL) lower limit on the charged Higgs boson mass of 80.0\GeV for the Type II scenario and of 72.5\GeV for the Type I scenario for ${m_{\PSA}} > 12\GeV$.
If the mass of the charged Higgs boson, ${m_{\PH^{+}}}$, is smaller than the mass difference between the top and the bottom quarks, the top quark can decay via  \tHpb (charge conjugate processes are always implied).
For values of $\tan\beta < 1$, the MSSM charged Higgs boson predominantly decays to a charm quark and a strange antiquark (\csbar).
In 2HDMs of Types I and Y~\cite{PhysRevD.80.015017}, the branching fraction $\mathcal{B}(\Hpcsbar)$ is larger than 10\% for any value of $\tan\beta$,
while in Types II and X it can reach $100\%$ for $\tan\beta$ $<$ 1.
In this study, we assume $\mathcal{B}(\Hpcsbar)$ to be $100\%$.

The presence of the \tHpb , \Hpcsbar decay mode alters the event yield of \ttbar pairs with hadronic jets in the final state, compared to the SM.
Upper limits on the branching fraction, $\mathcal{B}(\tHpb) <$ 10--20\%, have been set by the CDF~\cite{Aaltonen:2009ke} and D0~\cite{Abazov:2009aa} experiments at the Tevatron for $m_{\PH^{+}}$ between 80 and 155\GeV, assuming $\mathcal{B}(\Hpcsbar) = 100\%$.
Based on 4.7\fbinv of data recorded at a centre-of-mass energy of 7\TeV, the ATLAS Collaboration has set an upper limit on $\mathcal{B}(\tHpb)$ between 1 and 5\% for a charged Higgs boson mass in the range 90--150\GeV~\cite{Aad:2013hla}.

\begin{figure}[htp]
\centering
\includegraphics[width=0.47\textwidth]{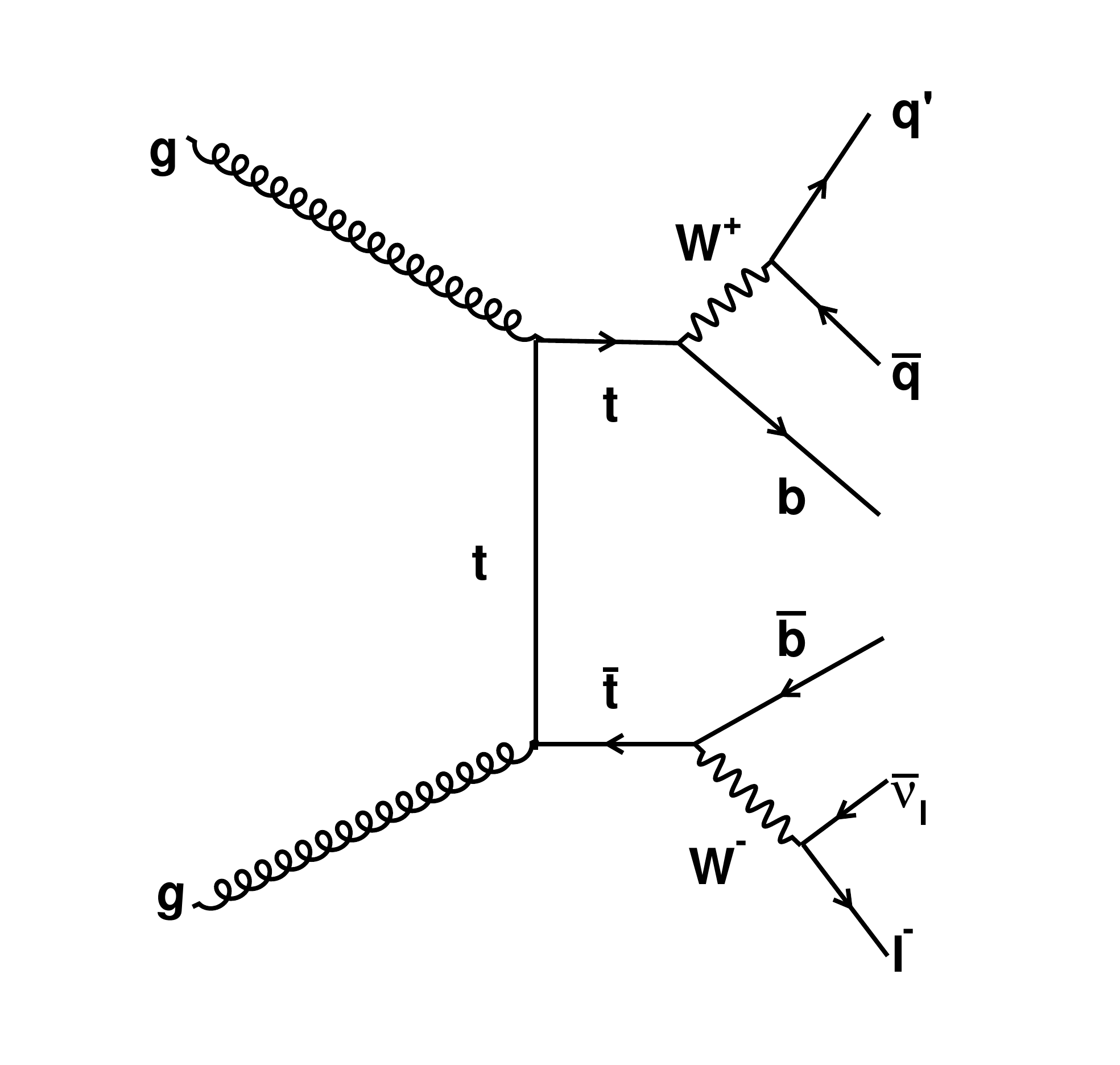} \hfill
\includegraphics[width=0.47\textwidth]{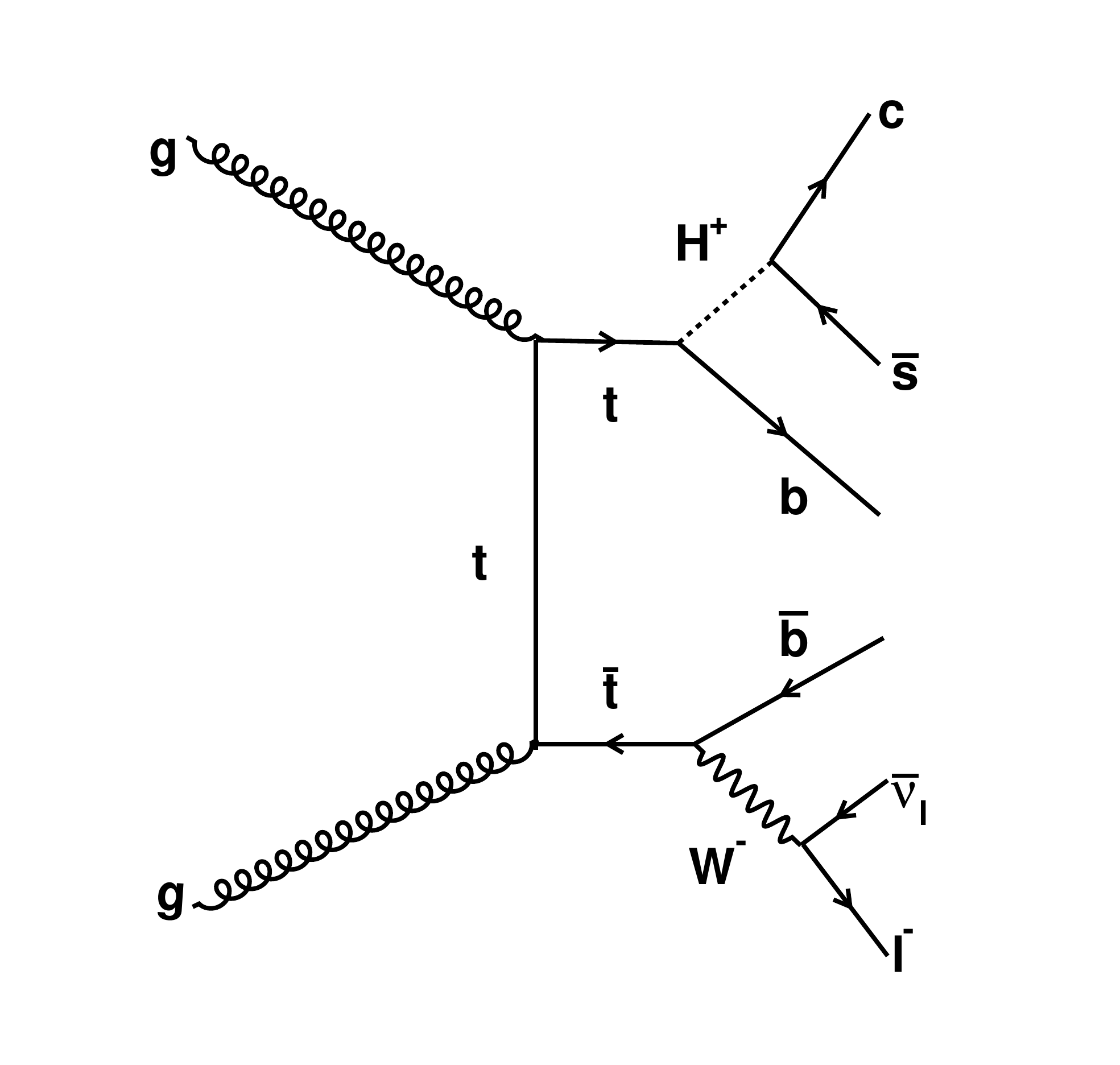}
\caption{Leading order Feynman diagram for \ttbar production at the LHC in the $\ell$+jets final state in the SM (left) and additional diagram for the model with a charged Higgs boson (right).
}
\label{fig:feynHiggs}
\end{figure}

In this paper, we report a model-independent search for a charged Higgs boson in the 90--160\GeV mass range using the final state ${\ttbar\to \cPqb\PH^{+}\cPaqb\PW^{-}}$, where the $\PW$ boson decays to a lepton (${\ell}=\Pe$ or $\Pgm$) and a neutrino, and the charged Higgs boson decays to \csbar.
The contribution of the process ${\ttbar\to \cPqb\PH^{+}\cPaqb\PH^{-}}$ is expected to be negligible in this $\ell$+jets final state.
Figure~\ref{fig:feynHiggs} shows the dominant Feynman diagrams for the final state both in the SM \ttbar process as well as for the model with a charged Higgs boson. We use a data sample recorded by the CMS experiment at the CERN LHC in pp collisions at $\sqrt{s} = 8\TeV$ corresponding to an integrated luminosity of 19.7\fbinv.

\section{The CMS detector, simulation and reconstruction}
A detailed description of the CMS detector together with a definition of the coordinate system used and the relevant kinematic variables can be found in Ref.~\cite{Chatrchyan:2008zzk}.
The central feature of CMS is a superconducting solenoid of 6\unit{m} diameter providing a magnetic field of 3.8\unit{T}.
A silicon pixel and strip tracker, a lead tungstate crystal electromagnetic calorimeter, and a brass-scintillator hadron calorimeter are located inside the solenoid.
Forward calorimeters extend the pseudorapidity~\cite{Chatrchyan:2008zzk} coverage provided by the barrel and endcap detectors.
The muon detection system is composed of drift tubes, cathode strip chambers, and resistive plate chambers, embedded in the steel flux-return yoke outside the solenoid.

The CMS first-level trigger system consists of custom hardware processors.
It uses information from the calorimeters and muon detector to select the interesting events.
The high-level trigger system, based on a computing farm, further reduces the event rate from around 100\unit{kHz} to less than 1\unit{kHz}, before data storage.

The analysis exploits event reconstruction based on the particle-flow (PF) algorithm~\cite{c:pft-09-001, CMS:2010byl}.
This algorithm reconstructs all stable particles in an event by combining information from all subdetectors.
The resulting list of particles is then used to reconstruct higher-level objects such as jets and missing transverse energy (\MET).
Muons are reconstructed by performing a simultaneous global track fit to hits in the silicon tracker and the muon detector~\cite{Chatrchyan:2012xi}.
Electrons are identified by combining information from clusters of energy deposits in the electromagnetic calorimeter and the hits in the tracker~\cite{Khachatryan:2015hwa}.
Jets are reconstructed using the anti-$\kt$ algorithm~\cite{Cacciari:2008gp} with a distance parameter of $0.5$.

Among the large number of pp interactions per LHC bunch crossing (``pileup'') we select the one having the maximum squared sum of the transverse momenta (\pt) of charged-particle tracks as the primary vertex.
On average, there were about $20$ pileup events in the 2012 data.
In order to suppress jets coming from pileup interactions, a jet identification criterion~\cite{CMS-PAS-JME-13-005} based on a multivariate analysis method is used.
We correct for the detector response to obtain a realistic jet energy scale.
The \MET~\cite{Chatrchyan:1361632} is defined as the magnitude of the vector sum of \pt of all reconstructed particles.

The method to identify jets from $\cPqb$ quark hadronization (called ``$\cPqb$ jets'') involves the use of secondary vertices together with track based lifetime information~\cite{Chatrchyan:2012jua,CMS-PAS-BTV-13-001} to provide an efficient discrimination between $\cPqb$ jets and jets from light quarks and gluons.
We choose a discriminator value that yields a misidentification probability for light-parton jets of approximately $1\%$ in the \pt range 80 to 120\GeV.
The corresponding $\cPqb$ tagging efficiency is ${\sim}70\%$ for jets with an average \pt of 80\GeV in \ttbar events.
The probability of misidentifying a $\cPqc$ jet as a $\cPqb$ jet is $\sim$20\%.

Background events from \ttbar decay processes (other than signal), $\PW$+jets and $\cPZ$+jets are generated with \MADGRAPH 5.1~\cite{Alwall:2011uj}, interfaced with \PYTHIA 6.4~\cite{Sjostrand:2006za}. The underlying event tuning Z2*~\cite{Chatrchyan:2011id} and CTEQ6M~\cite{Pumplin:2002vw} parton distribution function (PDF) set are used. The number of \ttbar events is estimated from the SM next-to-next-to-leading-order (NNLO) calculation~\cite{Czakon:2013goa} of the \ttbar production cross section, which is $252.9\pm 6.0$\,pb.
The single top quark processes are generated using {\POWHEG} 1.0~\cite{Nason:2004rx, Frixione:2007vw, Alioli:2010xd, Alioli:2009je, Re:2010bp}.
The expected contribution of the $\PW$+jets background is calculated at NNLO with \textsc{Fewz} 3.1~\cite{Li:2012wna}. The $\cPZ$+jets and single top quark events are also normalized to NNLO cross section calculations~\cite{Gavin:2010az,Kidonakis:2012db}.
The ${\ttbar\to \cPqb\PH^{+}\cPaqb\PW^{-}}$ ($\PH\PW$) signal sample is generated with \PYTHIA and normalized using the same production cross section as SM \ttbar. The diboson backgrounds ($\PW\PW$, $\PW\cPZ$, and $\cPZ\cPZ$) are generated with \PYTHIA and their cross sections are computed with \textsc{mcfm} 6.2~\cite{Campbell:2010ff}.

Generated events are processed through a full detector simulation based on {\GEANTfour}~\cite{Agostinelli:2002hh}, followed by a detailed trigger emulation and the CMS event reconstruction.
Minimum bias events are superimposed on the hard interactions to simulate pileup.
Simulated events are reweighted according to the pileup distribution observed in the data.

\section{Analysis}

\subsection{Event selection}

For the muon+jets final state, events are selected at the trigger level using an isolated single muon with $\pt > 24\GeV$ and $\abs{\eta}< 2.1$.
In the offline analysis, an event is selected if it has at least one reconstructed muon with $\pt > 25\GeV$ and $\abs{\eta} < 2.1$.
The muon is required to be isolated from the rest of the event activity by requiring the relative isolation $\mathrm{I}_\text{rel} < 0.12$, defined as
\begin{equation}
\label{eqn:pas_muoniso}
\mathrm{I}_\text{rel} = \frac{\mathrm{I}^\text{ch} + \max \left[ (\mathrm{I}^{\cPgg} + \mathrm{I}^\mathrm{nh} - 0.5\times \mathrm{I}^\text{ch}_\mathrm{PU}), 0 \right]}{\pt}.
\end{equation}

Here, $\mathrm{I}^\text{ch}$, $\mathrm{I}^{\cPgg}$, and $\mathrm{I}^\mathrm{nh}$ are the sum of transverse energies for charged hadrons, photons, and neutral hadrons, respectively, in a cone size of $\Delta R = \sqrt{\smash[b]{(\Delta\eta)^{2} + (\Delta\phi)^{2}}} = 0.4$ around the muon direction, and $\mathrm{I}^\text{ch}_\mathrm{PU}$ is the \pt sum of charged hadrons associated to all pileup vertices.
The latter term is used to estimate the contribution of neutral particles from the pileup events.
The factor $0.5$ takes into account the neutral-to-charged particle ratio.
The simulated events are reweighted in order to reproduce the muon trigger and selection efficiencies that are measured in data using a ``tag-and-probe'' technique~\cite{CMS:2011aa}.

For the electron+jets final state, events are selected with one isolated single-electron trigger with $\pt > 27\GeV$ and $\abs{\eta}< 2.5$; they are selected offline if the electron has $\pt > 30\GeV$ and $\abs{\eta}< 2.5$.
Other electron identification criteria are applied based on a multivariate analysis~\cite{Hocker:2007ht}. The electron should be isolated by requiring the relative isolation $\mathrm{I}^{\rho}_\text{rel} < 0.1$, given by
\begin{equation}
\label{eqn:eleiso}
\mathrm{I}^{\rho}_\text{rel} = \frac{\mathrm{I}^\text{ch} + \max \left[ (\mathrm{I}^{\cPgg} + \mathrm{I}^\mathrm{nh} - \rho {\rm A}_{\rm eff}), 0 \right]}{\pt},
\end{equation}
where $\mathrm{I}^\text{ch}$, $\mathrm{I}^{\cPgg}$, and $\mathrm{I}^\mathrm{nh}$ are calculated in a cone size of $\Delta {\rm R} = 0.3$ around the electron direction, $\rho$ is the energy density in the event that is used to estimate the average pileup contribution within the electron isolation cone, and ${\rm A}_{\rm eff}$ is a measure of the effective area subtended by the isolation cone.
Any event that has an additional muon or electron with $\pt > 10\GeV$ and $\abs{\eta}< 2.5$ passing a loose isolation (${<} 0.3$) is rejected.
The second-lepton veto rejects most of the events from $\cPZ$+jets and SM \ttbar decays, where both the $\PW$ bosons decay to leptons.

Events are required to have at least four jets with $\pt > 25\GeV$ and $\abs{\eta} < 2.5$, where two jets are expected to originate from top-quark decays and the other two from $\PW/\PH^{+}$ boson decays.
Since a neutrino is present in the signal final state, the events are required to have $\MET > 20\GeV$.
The \MET requirement suppresses the QCD multijet and $\cPZ(\ell^+\ell^-)$+jets backgrounds.
In these events the reconstructed \MET is expected to be small, arising mostly from the mismeasurement of energy in the calorimeters.
Compared to the dominant SM \ttbar background, the possible contribution from ${\ttbar{\rm V}\ ({\rm V}=\PW,\cPZ)}$ events is found to be negligible (less than $1\%$ of the total background).

In both signal and SM \ttbar events, there are two $\cPqb$ quarks in the final state that originate directly from the top quark decays.
Thus, we require the events to have at least two $\cPqb$ jets.
This requirement strongly suppresses the $\PW$+jets and QCD multijet backgrounds, where the $\cPqb$ jets come from the misidentification of light-quark including $\cPqc$ jets or gluon jets.
The simulated events are reweighted to reproduce the efficiencies measured in data in dedicated control regions~\cite{CMS-PAS-BTV-13-001}.

The \pt spectra of the top quark and antiquark in data are found to be softer than that predicted by the \MADGRAPH and \PYTHIA generators~\cite{Chatrchyan:2012saa}.
In order to account for this effect, the simulated \ttbar events are reweighted according to the generated \pt distribution of the top quarks and antiquarks. Event-by-event scale factors are derived based on the measurement of differential top-quark pair production cross
sections in the $\ell$+jets channel by CMS at $\sqrt{s} = 8\TeV$~\cite{Khachatryan:2015oqa}.
These factors are applied to the simulated SM \ttbar and signal samples before any event selection is required.

\subsection{Background estimation}
\label{s:pas_bkg_est}
Most of the backgrounds coming from \ttbar, $\PW$+jets, $\cPZ$+jets, single top quark and diboson processes are estimated from simulated samples normalized to NNLO predictions. As the QCD background is not well modeled by simulation, its contribution is estimated from data. A control region where the lepton candidate is non-isolated, given by $0.12  < \mathrm{I}_\text{rel} <  0.30$ for muons and $0.1 < \mathrm{I}^{\rho}_\text{rel} <  0.3$ for electrons, is used to estimate the normalization of the QCD multijet background.
After subtracting the expected contributions of other processes from data, the result is extrapolated to the signal region by using a scale factor determined from events with low \MET.
The shape of the QCD background distribution is evaluated from the sample of non-isolated leptons.

\begin{figure}[htp]
\centering
\includegraphics[width=0.60\textwidth]{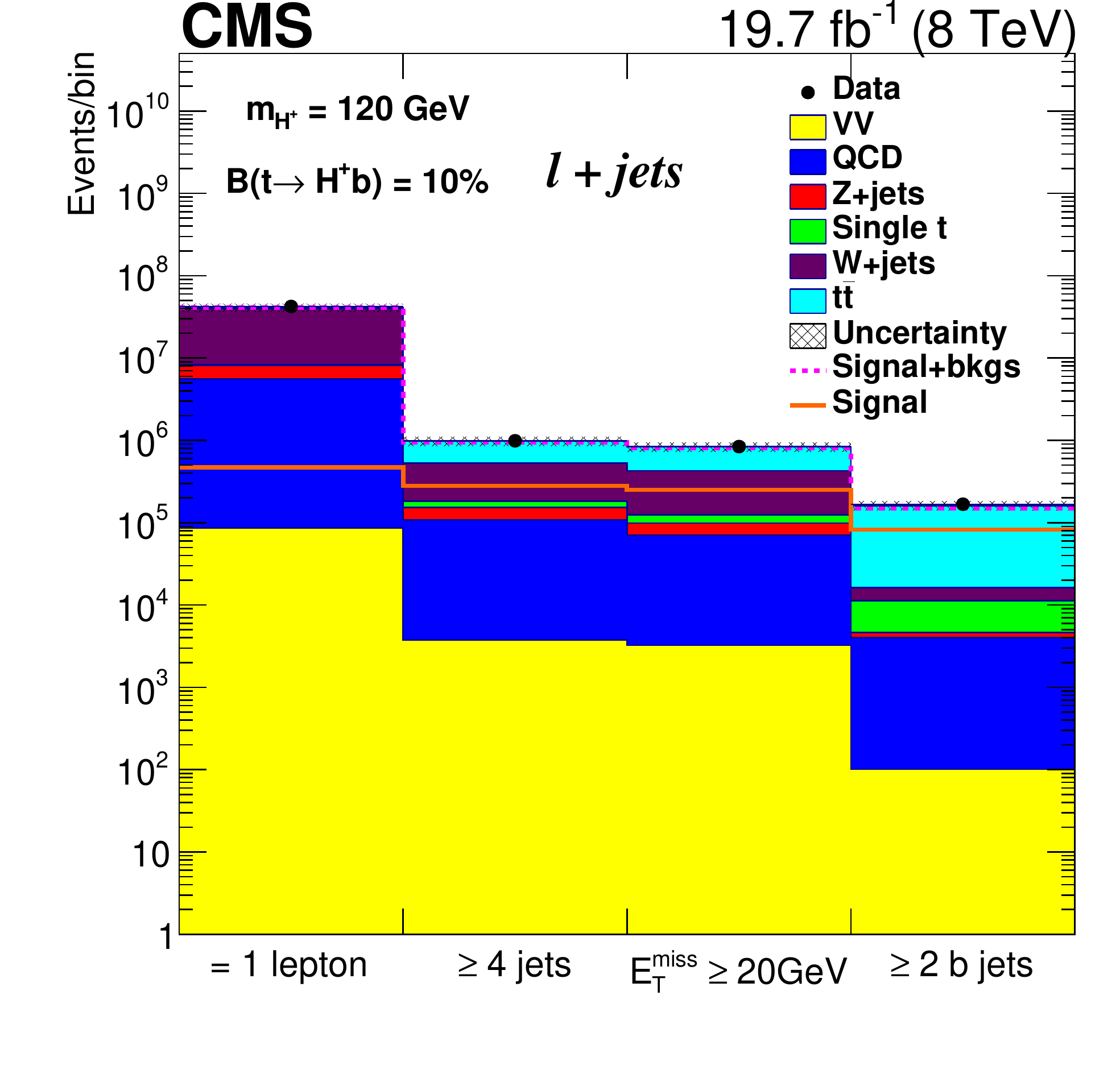}
\caption{Number of expected and observed data events after different selection requirements. ``Uncertainty'' in the legend here, and of other plots, includes statistical and systematic uncertainties.}
\label{fig:pas_cutflow}
\end{figure}

In Fig.~\ref{fig:pas_cutflow} we compare the event yields for various background samples and a signal sample, generated assuming ${m_{\PH^+}=120\,\GeV}$ and $\mathcal{B}(\tHpb)=10\%$, after each selection step.
At each step, the number of expected background events is found to match the data within uncertainties.
The dotted line in Fig.~\ref{fig:pas_cutflow} shows the total number of expected signal and background events in the presence of ${\PH^{+}}$ in the top quark decay. The total number of events including the ${\PH^{+}}$ signal is
\begin{equation}
{N_\text{total}}=(1-x)^{2}{N}_{\ttbar\to \cPqb\PW^{+}\cPaqb\PW^{-}}+2x(1-x){N}_{\PH\PW}+{ N_\text{other}},
\label{eqn:pas_Ntotal}
\end{equation}
where $x = \mathcal{B}(\tHpb)$ and $N_i$ is the number of expected events for the process $i$.
Based on simulations we have found that the expected contribution of the signal ${\ttbar\to \cPqb\PH^{+}\cPaqb\PH^{-}}$ component is negligible.

\subsection{Reconstruction of the \texorpdfstring{$\PW/\PH$}{W/H} mass}
\label{s:pas_kinfit}
A kinematic fit is used to fully reconstruct \ttbar events resulting in an improved mass resolution of the hadronically decaying boson.
The fit constraints the event to the hypothesis for the production of two top quarks.
As described above, one of the $\PW$ bosons from top quarks decays into a lepton-neutrino pair, while the other boson (a $\PW$ in SM \ttbar and the $\PH^{+}$ in the case of signal) decays into a quark-antiquark pair.
Since we are interested in reconstructing the $\PW/\PH^{+}$ boson mass, we relax the constraint on the light dijet mass to be consistent with the $\PW$ mass. On the other hand, both the top-quark masses are constrained to 172.5\GeV.
The detailed description of the algorithm and constraints on the fit are available in Ref.~\cite{Chatrchyan:2012cz}.

The kinematic fit receives the four-momenta of the lepton and all jets passing the selection requirements, \MET, and their respective resolutions.
The jet energy resolution (JER) in data varies between 5 to 20\% over the \pt range 30 to 1000\GeV. The jet energy in the simulation is thus smeared to appropriately reproduce the resolution measured in the data~\cite{Chatrchyan:2011ds}.

\begin{figure}[htbp]
\centering
\includegraphics[width=0.6\textwidth]{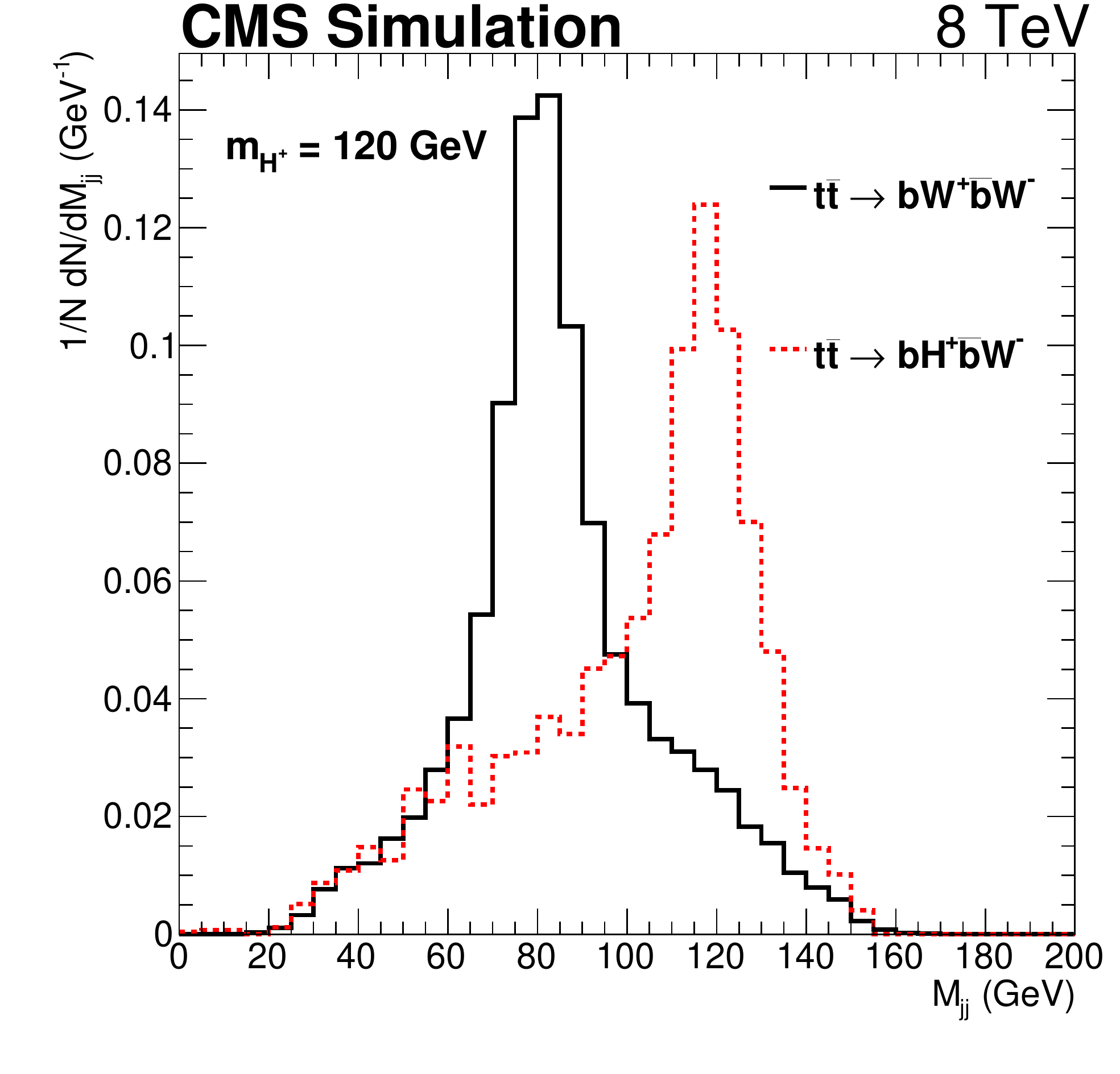}
\caption{Invariant mass distributions of the dijet system, assumed to come from \csbar hadronization, obtained with a kinematic fit after all selections. The solid black histogram represents the SM \ttbar events and the dashed red histogram denotes the same in the presence of a $\PH^{+}$ boson.
}
\label{fig:pas_w_mass}
\end{figure}

Only jets that pass the $\cPqb$ tagging requirement are considered as $\cPqb$-jet candidates in the \ttbar hypothesis, while all other jets are taken to be the light-quark candidates for hadronic boson decays.
For each event, the assignment that gives the maximum fit probability is retained.
The fit modifies the measured value of the jet \pt within its resolution to a value corresponding to the minimum $\chi^{2}$.
Figure~\ref{fig:pas_w_mass} shows the $\PW$ and $\PH^{+}$ boson mass distributions obtained from the kinematic fit after final event selections.
The kinematic fit significantly improves the dijet mass resolution, which is vital in separating the $\PH^{+}$ boson from the $\PW$ boson peak.

\begin{figure}[htbp]
\centering
\includegraphics[width=0.60\textwidth]{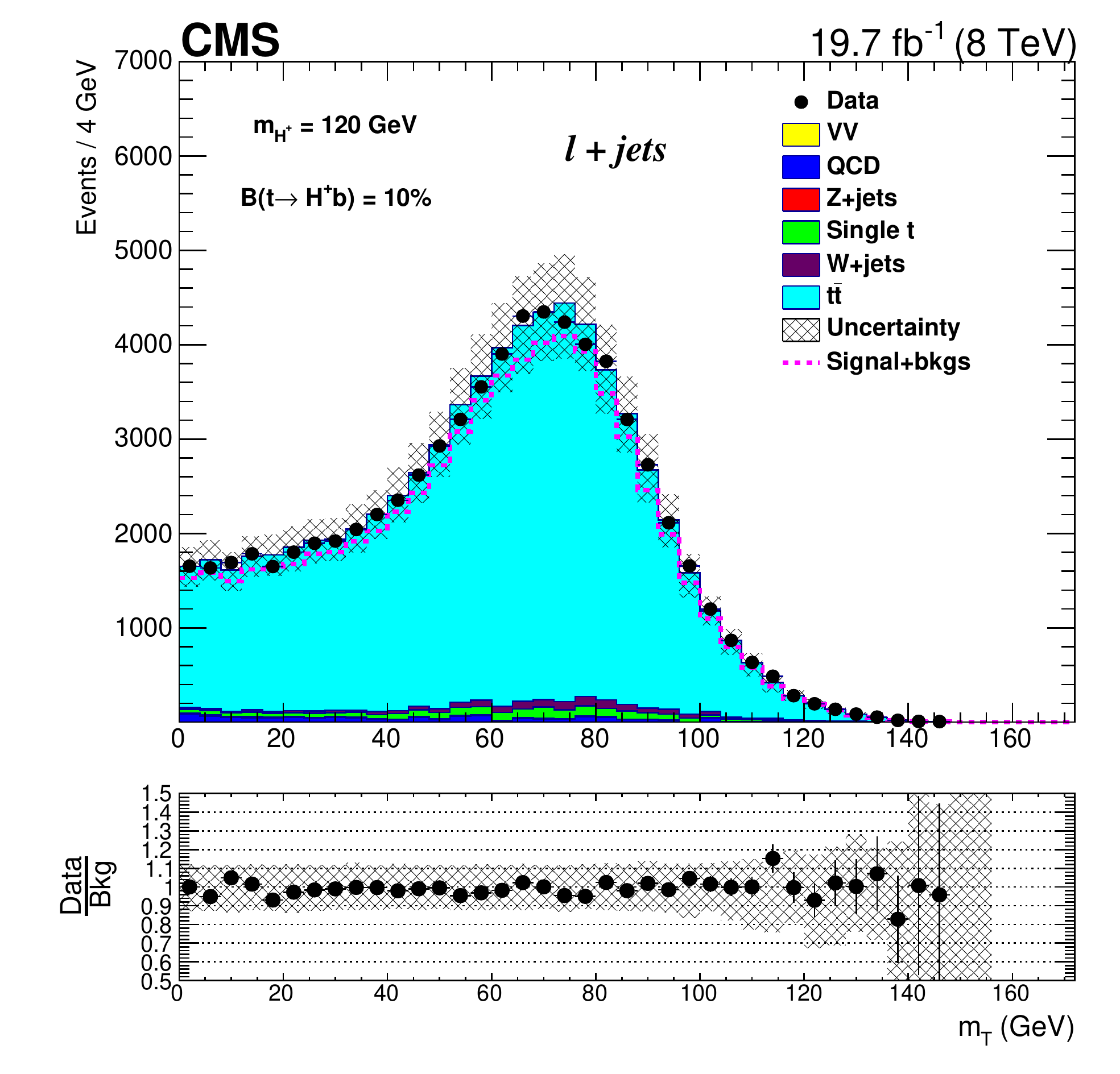}
\caption{Transverse mass distribution of the lepton plus $\MET$ system after all selections.}
\label{fig:wmt_kfit}
\end{figure}

As a control plot, Fig.~\ref{fig:wmt_kfit} shows the transverse mass ($m_{\mathrm{T}}$) of the system formed from the lepton candidate and \MET.
The dotted line represents the total number of expected signal and background events in the presence of $\PH^{+}$ obtained using Eq.~(\ref{eqn:pas_Ntotal}) for an assumed branching fraction of $\mathcal{B}(\tHpb) = 10\%$.
This line is below the total background expectation, because the reduction in the SM \ttbar yield is not fully compensated by the $\PH\PW$ component and we are missing the ${\ttbar\to \cPqb\PH^{+}\cPaqb\PH^{-}}$ component as we select the events with a high \pt lepton.
Other distributions such as \pt and $\eta$ of the lepton, jets, and $\cPqb$ jets as well as the jet multiplicity and $\cPqb$-jet multiplicity were studied. The $\chi^2$ distribution of the kinematic fit was also checked.
All these distributions show a good agreement between data and expected SM background.

\section{Systematic uncertainties}
\label{s:pas_systematics}
The following sources of systematic uncertainty are considered in this analysis.
\begin{itemize}

\item {{\bf Jet energy scale, resolution, and {\boldmath \MET} scale}: The uncertainty in the jet energy scale (JES) is the leading source of uncertainty in the analysis.
It is evaluated as a function of jet \pt and $\eta$ according to Ref.~\cite{Chatrchyan:2011ds}, and is then propagated to \MET.
The uncertainty in JES affects both the event yield and the shape of the dijet ($\PW$ or $\PH^{+}$) mass distribution.
To evaluate the uncertainty in the dijet mass distribution, the momenta of the jets are scaled according to the JES uncertainty by ${\pm}1\sigma$.
The scaled jet momenta are then passed on as inputs to the kinematic fit and the corresponding dijet mass is returned by the fit.
We take the difference in the dijet mass spectrum with respect to the nominal one as a shape uncertainty in the reference distribution used in the statistical analysis (described in Section~\ref{s:pas_results}).
In order to take the uncertainty due to the JER scale factor into account, two alternative dijet invariant mass distributions are obtained after smearing the jets with the JER scale factor varied by ${\pm}1\sigma$. The difference with respect to the nominal value is assigned as a shape uncertainty.
}

\item {{\bf {\boldmath $\cPqb$} tagging uncertainty}: The uncertainty in the $\cPqb$ tagging efficiency and misidentification probability is another leading source of uncertainty as the selection requires two $\cPqb$ jets.
The data-simulation scale factor and the corresponding uncertainty due to the $\cPqb$ tagging efficiency as well as the misidentification probability are taken from Ref.~\cite{CMS-PAS-BTV-13-001}.
The scale factor is applied to simulated events by randomly removing or promoting the events according to the scale factor in the $\cPqb$ tagging efficiency and misidentification probability. The uncertainty is estimated as the difference in the event yield when the scale factors are varied by their uncertainties. The data-simulation scale factor on the $\cPqc\to \cPqb$ misidentification probability is taken to be the same as that of the $\cPqb$ tagging efficiency and the error in the scale factor is taken as twice the corresponding uncertainty for the $\cPqb$ jets.
}

\item {{\bf Normalization uncertainty}: The error in the \ttbar production cross section, which is common for both SM \ttbar and signal events, is a leading source of uncertainty. We consider the uncertainties on the normalization of the $\PW$+jets and $\cPZ$+jets processes as fully correlated since their PDF uncertainties are known to be approximately 95\% correlated. The normalization uncertainties due to the single top quark and diboson processes are also considered.
}

\item {{\bf Lepton trigger, identification, and isolation  efficiency}: The uncertainty in the combined data-simulation scale factor for the muon trigger, identification, and isolation efficiencies is taken to be 3\%, as estimated using a tag-and-probe method. Similarly in the electron case the uncertainty on the associated combined data-simulation scale factor is taken to be 3\%.
}

\item{{\bf Uncertainty due to top quark {\boldmath $\pt$} reweighting}: As the top quark \pt reweighting is expected to change the dijet mass shape, the uncertainty corresponding to this reweighting is considered as a shape uncertainty~\cite{Khachatryan:2015oqa}.
}

\item {{\bf {\boldmath \ttbar} modeling uncertainty}: The uncertainty due to the variation of the renormalization and factorization scales used in the \ttbar simulation is estimated by simultaneously changing their common nominal value by factors of $0.5$ and $2$.
The nominal value is set to the momentum transfer ($Q$) in the hard process, given by $Q^{2}=m^{2}_{\cPqt}+\sum\pt^{2}$ in \MADGRAPH, where the sum is over all additional final state partons in the matrix element calculations.
An additional shape uncertainty is used to take into account the error due to matching thresholds used for interfacing the matrix elements generated with \MADGRAPH and \PYTHIA parton showering.
The thresholds are changed from the default value of 20\GeV to 10 and 40\GeV.
}

\item {{\bf Top mass uncertainty}: The uncertainty due to a possible variation of the top quark mass from its nominal value of 172.5\GeV used in the simulation is studied by changing the latter by ${\pm} 1\GeV$.
An additional shape uncertainty is used to take this effect into account.
}

\item {{\bf QCD normalization uncertainty}: As the QCD multijet contribution is obtained from data, we estimate the systematic uncertainty due to the error in the QCD scale factors from the non-isolated to isolated region by varying them by approximately 40\% and 60\% for the electron+jets and muon+jets channel, respectively. This is calculated using data as described in Section~\ref{s:pas_bkg_est}.}

\item {{\bf Size of the simulated samples}: Due to the limited size of some of the simulated samples, the bin-by-bin statistical uncertainties in the dijet mass distribution are taken into account for each of those samples.}

\item {{\bf Integrated luminosity uncertainty}: The uncertainty on the integrated luminosity measurement is estimated to be 2.6\%~\cite{CMS-PAS-LUM-13-001}.}

\end{itemize}

All systematic uncertainties considered for the muon+jets and electron+jets channel are listed in Tables~\ref{tab:pas_systematics_incl_mu} and \ref{tab:pas_systematics_incl_ele}, respectively.

\begin{table}[htbp]
\centering
\topcaption{Systematic uncertainties (in percent) for the yield of signal and background processes after all selections in the muon+jets channel.}
\label{tab:pas_systematics_incl_mu}
\renewcommand{\arraystretch}{1.1}
\begin{tabular}{l.......}
\cline{2-8}
 \multicolumn{1}{c}{}
 & \multicolumn{1}{c}{$\PH\PW$} & \multicolumn{1}{c}{${\ttbar}_{\Pgm + {\rm jets}}$} & \multicolumn{1}{c}{$\PW$+jets} & \multicolumn{1}{c}{$\cPZ$+jets} & \multicolumn{1}{c}{Single $\cPqt$} & \multicolumn{1}{c}{Diboson} & \multicolumn{1}{c}{QCD} \\
\hline
 JES+JER+\MET & 6.0 & 3.2 & 24.9&19.6&6.4&11.5& \multicolumn{1}{c}{\NA} \\
$\cPqb$ tagging &5.6 &4.3& \multicolumn{1}{c}{\NA}& \multicolumn{1}{c}{\NA}&5.3& \multicolumn{1}{c}{\NA}& \multicolumn{1}{c}{\NA} \\
$\mathrm{Jet} \to \cPqb$ misidentification& \multicolumn{1}{c}{\NA}&  \multicolumn{1}{c}{\NA} &5.1&3.1&  \multicolumn{1}{c}{\NA} &3.7& \multicolumn{1}{c}{\NA} \\
Lepton selection &3.0&3.0&3.0&3.0&3.0&3.0& \multicolumn{1}{c}{\NA} \\
Normalization &6.0&6.0&5.0&4.0&6.0&10.0&60.0 \\
Simulation statistics&2.1&0.4&4.4&3.5&2.0&5.8&17.5 \\
Top quark \pt reweighting&3.6&1.3& \multicolumn{1}{c}{\NA}& \multicolumn{1}{c}{\NA}& \multicolumn{1}{c}{\NA}& \multicolumn{1}{c}{\NA}& \multicolumn{1}{c}{\NA}\\
Integrated luminosity &2.6&2.6&2.6&2.6&2.6&2.6& \multicolumn{1}{c}{\NA} \\
\hline
\end{tabular}
\end{table}

\begin{table}[htbp]
\centering
\topcaption{Systematic uncertainties (in percent) for the yield of signal and background processes after all selections in the electron+jets channel.}
\label{tab:pas_systematics_incl_ele}
\renewcommand{\arraystretch}{1.1}
\begin{tabular}{l.......}
\cline{2-8}
 \multicolumn{1}{c}{}
 & \multicolumn{1}{c}{$\PH\PW$} & \multicolumn{1}{c}{${\ttbar}_{\Pe + {\rm jets}}$} & \multicolumn{1}{c}{$\PW$+jets} & \multicolumn{1}{c}{$\cPZ$+jets} & \multicolumn{1}{c}{Single $\cPqt$} & \multicolumn{1}{c}{Diboson} & \multicolumn{1}{c}{QCD} \\
\hline
 JES+JER+\MET &4.6 & 4.0 & 18.3&17.5&6.0&12.5&\multicolumn{1}{c}{\NA} \\
$\cPqb$ tagging &5.6&4.3&\multicolumn{1}{c}{\NA}&\multicolumn{1}{c}{\NA}&5.9&\multicolumn{1}{c}{\NA}&\multicolumn{1}{c}{\NA} \\
$\mathrm{Jet} \to \cPqb$ misidentification&\multicolumn{1}{c}{\NA}& \multicolumn{1}{c}{\NA} &4.8&4.5& \multicolumn{1}{c}{\NA} &7.5&\multicolumn{1}{c}{\NA} \\
Lepton selection &3.0&3.0&3.0&3.0&3.0&3.0&\multicolumn{1}{c}{\NA} \\
Normalization &6.0&6.0&5.0&4.0&6.0&10.0&40.0 \\
Simulation statistics&2.5&0.5&5.4&5.1&2.5&6.3&13.8 \\
Top quark \pt reweighting&3.8&2.2&\multicolumn{1}{c}{\NA}&\multicolumn{1}{c}{\NA}&\multicolumn{1}{c}{\NA}&\multicolumn{1}{c}{\NA}&\multicolumn{1}{c}{\NA}\\
Integrated luminosity &2.6&2.6&2.6&2.6&2.6&2.6&\multicolumn{1}{c}{\NA} \\
\hline
\end{tabular}
\end{table}

\section{Results}
\label{s:pas_results}
\begin{figure}[htbp]
\centering
\includegraphics[width=0.45\textwidth]{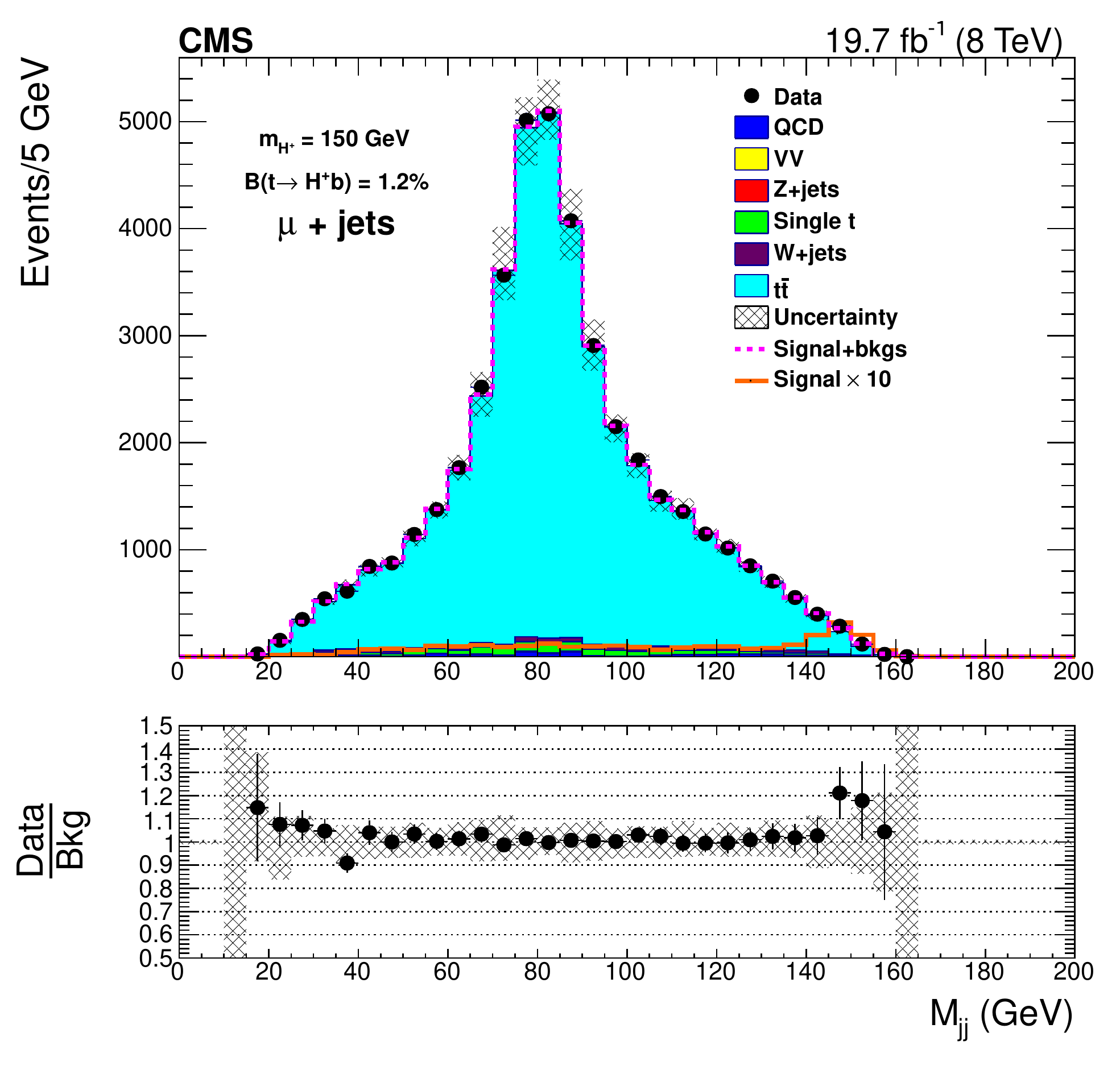} \hfill
\includegraphics[width=0.45\textwidth]{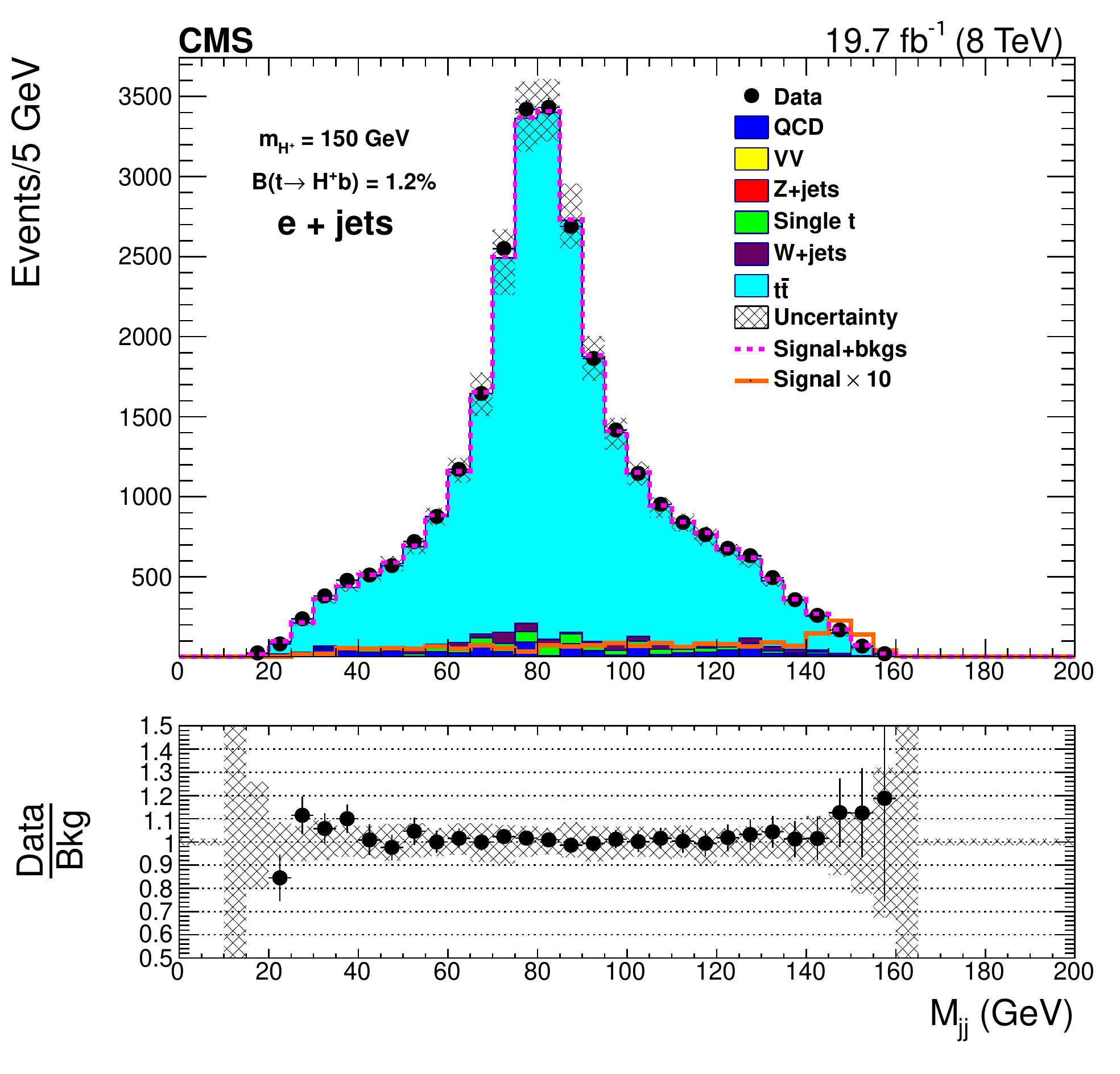}
\caption{Dijet mass distributions of the hadronically decaying boson after all selections, using background templates and constrained uncertainties obtained from the maximum likelihood fit, for the muon+jets (left) and electron+jets (right) channel. The dotted line represents the expected yield in the presence of signal.}
\label{fig:pas_post_fit_mjj_final_incl}
\end{figure}

The event yields after all selections are listed in Table~\ref{tab:pas_final_cutflowtable_incl} along with their combined statistical and systematic uncertainties. The number of signal events from the $\PH\PW$ process is also listed for $\mathcal{B} (\tHpb) = 10\%$.
The signal event yield is obtained using the SM \ttbar cross section. The total number of expected background events matches well the number of observed data events within uncertainties. The dijet mass distribution after all selections is shown in Fig.~\ref{fig:pas_post_fit_mjj_final_incl}.
The dotted line represents the expected distribution of signal and background events for $\mathcal{B} (\tHpb) = 10\%$.
Again, the data are in agreement with the SM background expectation.
An upper limit is obtained on $\mathcal{B}(\tHpb)$ as discussed later in this Section.
\begin{table}[htbp]
\centering
\topcaption{The number of expected signal and background events in 19.7\fbinv of data, along with their combined statistical and systematic uncertainties.}
\label{tab:pas_final_cutflowtable_incl}
\renewcommand{\arraystretch}{1.1}
\begin{tabular}{lx{7}x{8}}
\hline
\multicolumn{1}{l}{Process} & \multicolumn{1}{c}{muon+jets channel} & \multicolumn{1}{c}{electron+jets channel} \\
\hline
\multicolumn{1}{l}{ $\PH\PW$, $m_{\PH^{+}}=120\GeV$ } &
\multicolumn{1}{c}{\multirow{2}{*}{ $4420 \, \pm \, 580$ }} & \multicolumn{1}{c}{\multirow{2}{*}{ $2943 \, \pm \, 371$ }}\\
\multicolumn{1}{l}{ ${\mathcal{B}}(\tHpb) = 10\%$ } & & \\
\hline
SM \ttbar & 41712 , 4735 & 25884 , 3009 \\
$\PW$+jets & 755 , 199 & 500 , 101 \\
$\cPZ$+jets & 91 , 19 & 83 , 16 \\
QCD & 381 , 67 & 655 , 91 \\
Single $\cPqt$ & 1096 , 114 & 687 , 73 \\
Diboson & 15 , 3 & 12 , 2 \\
\hline
Total background & 44050 , 4741 & 27820 , 3013 \\
\hline
Data  & \multicolumn{1}{c}{42785} & \multicolumn{1}{c}{28447} \\
\hline
\end{tabular}
\end{table}

Assuming that any excess or deficit of events in data, when compared with the expected background contribution, is due to the \tHpb, \Hpcsbar decay, the difference $\Delta N$ between the observed number of data events and the predicted background contribution is given as a function of $x=\mathcal{B}(\tHpb)$ via the following relation:
\begin{equation}
\Delta N = N^\mathrm{BSM}_{\ttbar}-N^\mathrm{SM}_{\ttbar} = 2x(1-x)N^{\PH\PW} + [(1 - x)^{2} - 1]N^\mathrm{SM}_{\ttbar}.
\label{eqn:pas_deltaN}
\end{equation}

Here, $N^{\PH\PW}$ is estimated from simulation forcing the first top quark to decay to $\cPqb\PH^{+}$ and the second to $\cPaqb\PW^{-}$, and $N^\mathrm{SM}_{\ttbar}$ is also calculated from simulation, as given by the \ttbar background in Table~\ref{tab:pas_final_cutflowtable_incl}.
Eq.~(\ref{eqn:pas_deltaN}) does not depend on any MSSM parameters. Therefore, the obtained limit in the absence of a significant excess or deficit of events is model-independent.

\begin{figure}[htbp]
\centering
\includegraphics[width=0.60\textwidth]{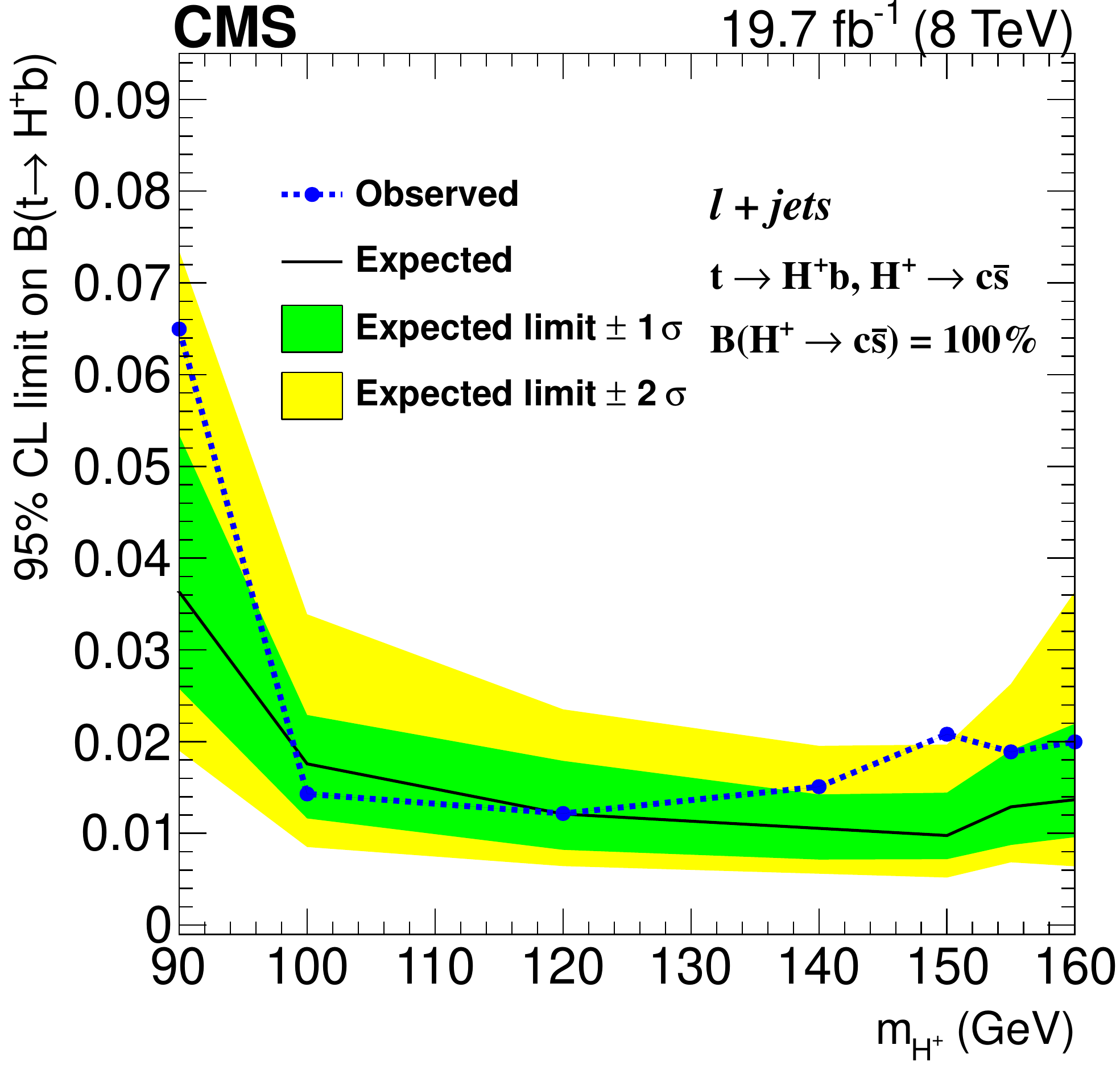}
\caption{Exclusion limit on the branching fraction $\mathcal{B}(\tHpb)$ as a function of $m_{\PH^+}$ assuming $\mathcal{B}(\Hpcsbar)=100\%$.}
\label{fig:pas_limit_incl}
\end{figure}

\begin{figure}[htbp]
\centering
\includegraphics[width=0.45\textwidth]{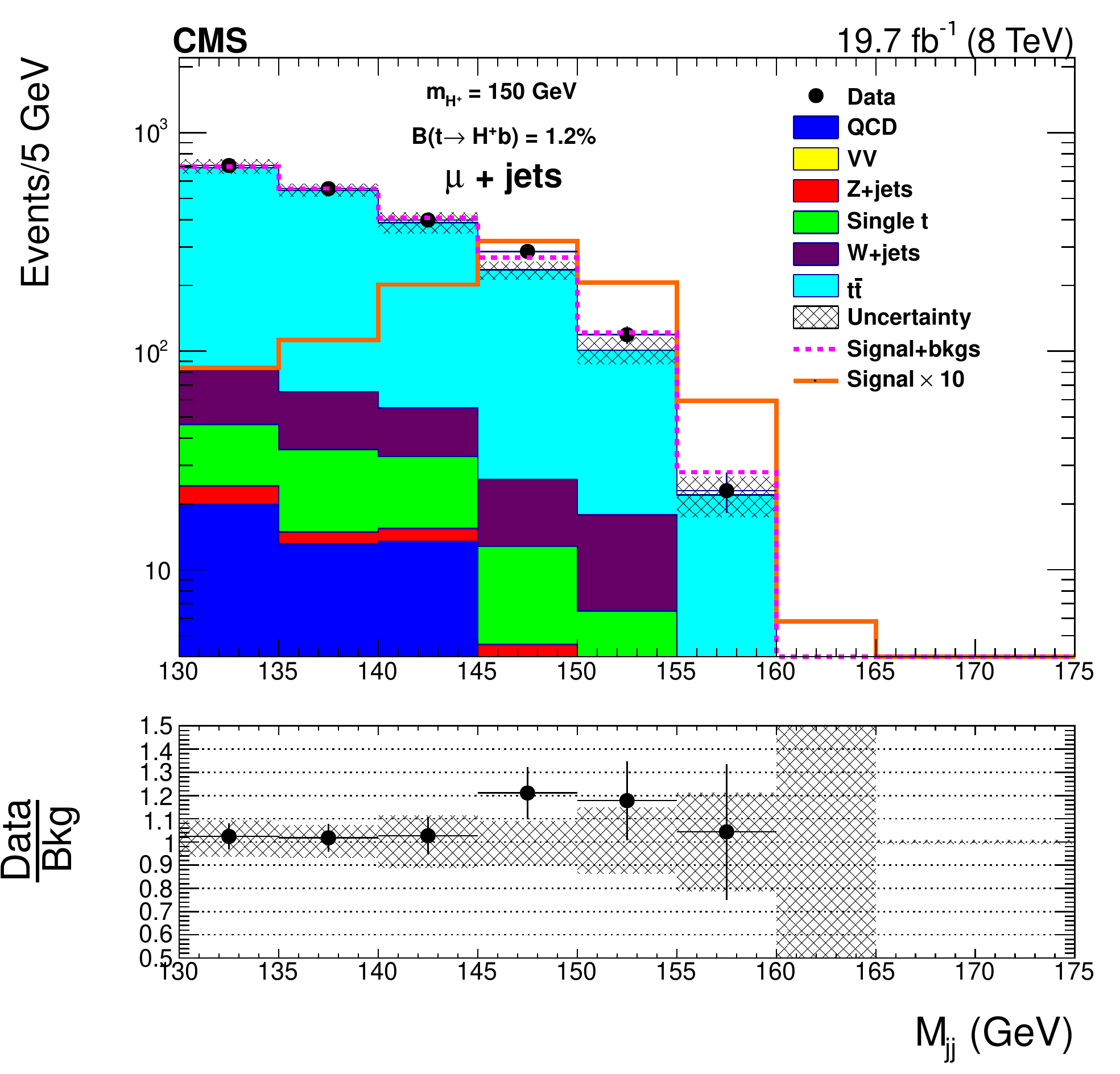} \hfill
\includegraphics[width=0.45\textwidth]{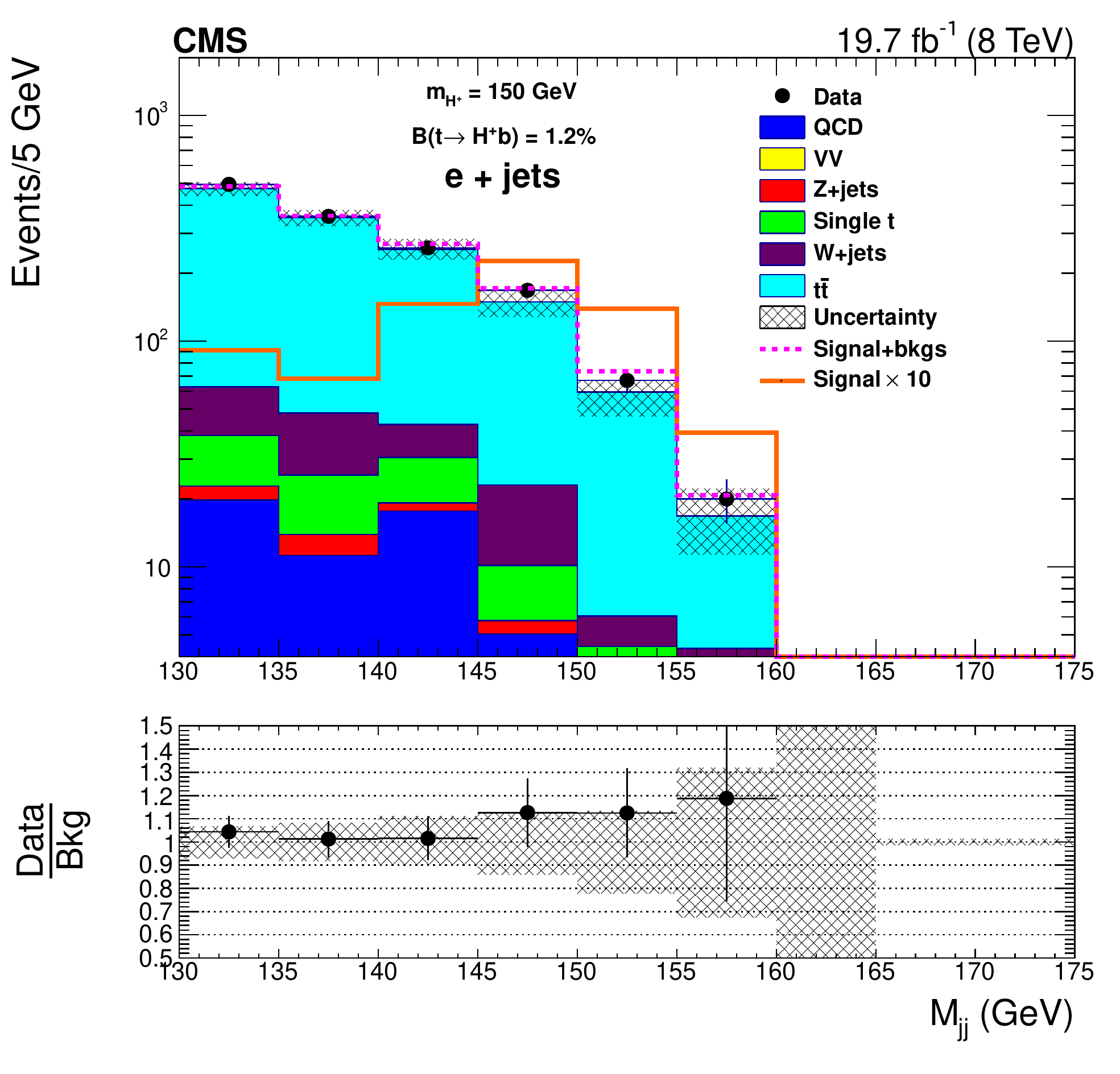}
\caption{An expanded view of the dijet mass distribution of the hadronically decaying boson after all selections, using background templates and constrained uncertainties obtained from the maximum likelihood fit, for the muon+jets (left) and electron+jets (right) channel. The dotted line represents the expected yield in the presence of signal.}
\label{fig:pas_post_fit_mjj_final_incl_zoom}
\end{figure}

Based on the $\rm CL_S$ method~\cite{CLs,Junk:1999kv}, we perform a binned maximum-likelihood fit to the dijet mass distributions shown in Fig.~\ref{fig:pas_post_fit_mjj_final_incl} in order to search for a possible signal.
The background and signal uncertainties described in Section~\ref{s:pas_systematics} are modeled with log-normal probability distribution functions.
These uncertainties are represented by nuisance parameters that are varied in the fit.
Correlations of all possible uncertainties between signal and backgrounds as well as among the two channels are taken into account.
An upper limit at the 95\% \CL is set on $\mathcal{B}(\tHpb)$ using Eq.~(\ref{eqn:pas_deltaN}).
Both the expected and observed limit as a function of $m_{\PH^{+}}$ are shown in Fig.~\ref{fig:pas_limit_incl}, while Table~\ref{tab:pas_limit_table} provides their numerical values.
The expected upper limit ranges between $1.0$ and $3.6\%$ for the mass range probed.
The observed limit agrees with the expected one within two standard deviations ($\sigma$), except for the region around 150\GeV where we see some excess.
To better understand this excess, in Fig.~\ref{fig:pas_post_fit_mjj_final_incl_zoom} we present an expanded view of the dijet mass distribution for the muon+jets and electron+jets channel.
We find the data points to be consistent with the signal-plus-background hypothesis for a charged Higgs boson mass $m_{\PH^{+}}=150\GeV$ for a best-fit branching fraction value $(1.2\pm0.2)\%$.
The quoted uncertainty here includes both statistical and systematic errors.
The local observed significance is $2.4\sigma$, which becomes $1.5\sigma$ after incorporating the look-elsewhere effect~\cite{CMS-NOTE-2011-005}, calculated over the mass region probed in a finer binning of 1\GeV.

\begin{table}[htbp]
\centering
\topcaption{Expected and observed limits on $\mathcal{B}(\tHpb)$ (in percent) at 95\% \CL in the mass range of 90 to 160\GeV.}
\label{tab:pas_limit_table}
\begin{tabular}{c|ccccc|c}
\hline
\multicolumn{7}{c}{95\% \CL upper limit on $\mathcal{B}(\tHpb)$ in percent }\\
\hline
 $m_{\PH^{+}}$   & \multicolumn{5}{c|}{Expected limit}                   &
 \multirow{2}{*}{ Observed limit } \\
 (\GeVns)   & ${-}2\sigma$  & ${-}1\sigma$ & median & ${+}1\sigma$ & ${+}2\sigma$  &
 \\
\hline
90 & 1.9 & 2.6 & 3.6 & 5.3 & 7.3&6.5\\
100 & 0.9 & 1.2 & 1.8 & 2.3 & 3.4&1.4\\
120 & 0.6 & 0.8 & 1.2 & 1.8 & 2.4&1.2\\
140 & 0.6 & 0.7 & 1.1 & 1.4 & 2.0&1.5\\
150 & 0.5 & 0.7 & 1.0 & 1.4 & 2.0&2.1\\
155 & 0.7 & 0.9 & 1.3 & 1.9 & 2.6&1.9\\
160 & 0.6 & 1.0 & 1.4 & 2.2 & 3.6&2.0\\
\hline
\end{tabular}
\end{table}

\section{Summary}
A search has been performed for a light charged Higgs boson produced in the top quark decay, subsequently decaying into a charm quark and a strange antiquark.
The data sample used in the analysis corresponds to an integrated luminosity of 19.7\fbinv recorded by the CMS experiment at $\sqrt{s} = 8\TeV$ in $\Pp\Pp$ collisions.
After analyzing the dijet mass distribution of the \Hpcsbar candidate events that comprise an isolated lepton, at least four hadronic jets, two of which are identified as $\cPqb$ jets, and large missing transverse energy, we have set model-independent upper limits on the branching fraction $\mathcal{B}(\tHpb)$ assuming $\mathcal{B}(\Hpcsbar)=100\%$. The 95\% confidence level upper limits are in the range 1.2--6.5\% for a charged Higgs boson mass between 90 and 160\GeV.

\begin{acknowledgments}
\hyphenation{Bundes-ministerium Forschungs-gemeinschaft Forschungs-zentren} We congratulate our colleagues in the CERN accelerator departments for the excellent performance of the LHC and thank the technical and administrative staffs at CERN and at other CMS institutes for their contributions to the success of the CMS effort. In addition, we gratefully acknowledge the computing centres and personnel of the Worldwide LHC Computing Grid for delivering so effectively the computing infrastructure essential to our analyses. Finally, we acknowledge the enduring support for the construction and operation of the LHC and the CMS detector provided by the following funding agencies: the Austrian Federal Ministry of Science, Research and Economy and the Austrian Science Fund; the Belgian Fonds de la Recherche Scientifique, and Fonds voor Wetenschappelijk Onderzoek; the Brazilian Funding Agencies (CNPq, CAPES, FAPERJ, and FAPESP); the Bulgarian Ministry of Education and Science; CERN; the Chinese Academy of Sciences, Ministry of Science and Technology, and National Natural Science Foundation of China; the Colombian Funding Agency (COLCIENCIAS); the Croatian Ministry of Science, Education and Sport, and the Croatian Science Foundation; the Research Promotion Foundation, Cyprus; the Ministry of Education and Research, Estonian Research Council via IUT23-4 and IUT23-6 and European Regional Development Fund, Estonia; the Academy of Finland, Finnish Ministry of Education and Culture, and Helsinki Institute of Physics; the Institut National de Physique Nucl\'eaire et de Physique des Particules~/~CNRS, and Commissariat \`a l'\'Energie Atomique et aux \'Energies Alternatives~/~CEA, France; the Bundesministerium f\"ur Bildung und Forschung, Deutsche Forschungsgemeinschaft, and Helmholtz-Gemeinschaft Deutscher Forschungszentren, Germany; the General Secretariat for Research and Technology, Greece; the National Scientific Research Foundation, and National Innovation Office, Hungary; the Department of Atomic Energy and the Department of Science and Technology, India; the Institute for Studies in Theoretical Physics and Mathematics, Iran; the Science Foundation, Ireland; the Istituto Nazionale di Fisica Nucleare, Italy; the Ministry of Science, ICT and Future Planning, and National Research Foundation (NRF), Republic of Korea; the Lithuanian Academy of Sciences; the Ministry of Education, and University of Malaya (Malaysia); the Mexican Funding Agencies (CINVESTAV, CONACYT, SEP, and UASLP-FAI); the Ministry of Business, Innovation and Employment, New Zealand; the Pakistan Atomic Energy Commission; the Ministry of Science and Higher Education and the National Science Centre, Poland; the Funda\c{c}\~ao para a Ci\^encia e a Tecnologia, Portugal; JINR, Dubna; the Ministry of Education and Science of the Russian Federation, the Federal Agency of Atomic Energy of the Russian Federation, Russian Academy of Sciences, and the Russian Foundation for Basic Research; the Ministry of Education, Science and Technological Development of Serbia; the Secretar\'{\i}a de Estado de Investigaci\'on, Desarrollo e Innovaci\'on and Programa Consolider-Ingenio 2010, Spain; the Swiss Funding Agencies (ETH Board, ETH Zurich, PSI, SNF, UniZH, Canton Zurich, and SER); the Ministry of Science and Technology, Taipei; the Thailand Center of Excellence in Physics, the Institute for the Promotion of Teaching Science and Technology of Thailand, Special Task Force for Activating Research and the National Science and Technology Development Agency of Thailand; the Scientific and Technical Research Council of Turkey, and Turkish Atomic Energy Authority; the National Academy of Sciences of Ukraine, and State Fund for Fundamental Researches, Ukraine; the Science and Technology Facilities Council, UK; the US Department of Energy, and the US National Science Foundation.

Individuals have received support from the Marie-Curie programme and the European Research Council and EPLANET (European Union); the Leventis Foundation; the A. P. Sloan Foundation; the Alexander von Humboldt Foundation; the Belgian Federal Science Policy Office; the Fonds pour la Formation \`a la Recherche dans l'Industrie et dans l'Agriculture (FRIA-Belgium); the Agentschap voor Innovatie door Wetenschap en Technologie (IWT-Belgium); the Ministry of Education, Youth and Sports (MEYS) of the Czech Republic; the Council of Science and Industrial Research, India; the HOMING PLUS programme of the Foundation for Polish Science, cofinanced from European Union, Regional Development Fund; the OPUS programme of the National Science Center (Poland); the Compagnia di San Paolo (Torino); the Consorzio per la Fisica (Trieste); MIUR project 20108T4XTM (Italy); the Thalis and Aristeia programmes cofinanced by EU-ESF and the Greek NSRF; the National Priorities Research Program by Qatar National Research Fund; the Rachadapisek Sompot Fund for Postdoctoral Fellowship, Chulalongkorn University (Thailand); and the Welch Foundation, contract C-1845.
\end{acknowledgments}
\bibliography{auto_generated}

\cleardoublepage \appendix\section{The CMS Collaboration \label{app:collab}}\begin{sloppypar}\hyphenpenalty=5000\widowpenalty=500\clubpenalty=5000\textbf{Yerevan Physics Institute,  Yerevan,  Armenia}\\*[0pt]
V.~Khachatryan, A.M.~Sirunyan, A.~Tumasyan
\vskip\cmsinstskip
\textbf{Institut f\"{u}r Hochenergiephysik der OeAW,  Wien,  Austria}\\*[0pt]
W.~Adam, E.~Asilar, T.~Bergauer, J.~Brandstetter, E.~Brondolin, M.~Dragicevic, J.~Er\"{o}, M.~Flechl, M.~Friedl, R.~Fr\"{u}hwirth\cmsAuthorMark{1}, V.M.~Ghete, C.~Hartl, N.~H\"{o}rmann, J.~Hrubec, M.~Jeitler\cmsAuthorMark{1}, V.~Kn\"{u}nz, A.~K\"{o}nig, M.~Krammer\cmsAuthorMark{1}, I.~Kr\"{a}tschmer, D.~Liko, T.~Matsushita, I.~Mikulec, D.~Rabady\cmsAuthorMark{2}, B.~Rahbaran, H.~Rohringer, J.~Schieck\cmsAuthorMark{1}, R.~Sch\"{o}fbeck, J.~Strauss, W.~Treberer-Treberspurg, W.~Waltenberger, C.-E.~Wulz\cmsAuthorMark{1}
% ...
\vskip\cmsinstskip
% ...
1:~~Also at Vienna University of Technology, Vienna, Austria\\
2:~~Also at CERN, European Organization for Nuclear Research, Geneva, Switzerland\\
% ...

\end{sloppypar}
\end{document}